\documentclass[preprint,12pt]{elsarticle}

\usepackage{graphicx}

\usepackage{amssymb}
\usepackage{amsthm}
\usepackage{amsmath}
\usepackage{subfigure}

\journal{Physica A}

\begin{document}

\begin{frontmatter}



\title{Phase Diagram and Thermodynamic and Dynamic Anomalies in a Pure Repulsive Model}


\author[label1]{Andressa A. Bertolazzo}
\author[label1]{Marcia C. Barbosa}

\address[label1]{Instituto de F\'{\i}sica, Universidade Federal do Rio Grande do Sul, \\
Caixa Postal 15051, 91501-970, Porto Alegre, RS, Brazil}

\begin{abstract}
Using Monte Carlo simulations a lattice gas model with only repulsive interactions was checked for the presence of anomalies. We show that this system exhibits the density (temperature of maximum density - TMD) and diffusion anomalies as present in liquid water. These anomalous behavior exist in the region of the chemical potential {\it vs} temperature phase diagram where two structured phases are present. A fragile-to-strong dynamic transition is also observed in the vicinity of the TMD line.
\end{abstract}


\end{frontmatter}



\section{Introduction}
\label{introduction}
Water is one of the most important liquids in the world. Its relevance to life and to the industrial processes are a consequence of many of its anomalous properties. One example is the density anomaly. While for most liquids density decreases monotonically as temperature increases at constant pressure, this is not the case of liquid water. For high temperatures the density decreases with the increase of temperature, but at approximately $4^oC$  at $1$ atm density reaches a maximum and decreases monotonically \cite{Ke67}. This water property is responsible for keeping the water liquid inside lakes and rivers at subzero temperatures while the surface is frozen. This is one of the 70 known water anomalous properties \cite{chaplin}. Another peculiar water behavior is that it can diffuse faster in a more dense state. For normal liquids we expect that the diffusion coefficient, $D$, at constant temperature increases with the decrease of pressure, since mobility is enhanced in a less dense medium. However, for water there is a range of pressures in which diffusion exhibits a non monotonic behavior with pressure and it increases as water gets more dense \cite{physicaA_2002}. \\

The diffusion coefficient of water exhibits another anomalous behavior: it changes from a non-Arrhenius diffusion to an Arrhenius diffusion when temperature is decreased at constant pressure in the region of pressure {\it vs} temperature phase diagram where the response functions have a large increase. Water is fragile (non-Arrhenius) at room and moderately supercooled temperatures \cite{An76}, but has been shown to be a strong liquid (with Arrhenius behavior) by dielectric relaxation measurements near the glass transition temperature \cite{fstran} for confined water. Experiments using NMR and quasi-elastic neutron scattering  \cite{fst2} and with nuclear magnetic resonance \cite{fst3} confirm the fragile-to-strong transition in the ``no-man's land''. This is the region of the pressure {\it vs} temperature phase diagram where no liquid phase is found since homogeneous nucleation takes place. However, crystallization can be avoid in confined systems. Recently, the existence of a fragile-to-strong transition is associated with the presence of a liquid-liquid transition  as follows. Experiments and simulations show that a fragile-to-strong transition occurs when the continuation of the liquid-liquid phase transition, the so called Widom line, is crossed at constant pressure \cite{fst_stanley}. \\

Water can form hydrogen bonds and this interaction between molecules has been considered the main mechanism for anomalies \cite{Be33}. Due to the hydrogen bonds water molecules organizes themselves into arrangements of four bonded molecules, the tetramers. They can form two types of structures, one more rigid and open in which all molecules are bonded and another more malleable and close in which only part of the molecules are bonded. These two structures are present in the formation of hydrogen bond network in percentages that vary. While the open structure occupies less volume and requires less pressure, the closed structure has lower energy. This competition is responsible for the density and diffusion anomaly \cite{debenedeti_nature_2010,Allan_JCP, jagla, stell} and to the reentrant multiple solid phases \cite{ryzhov1,ryzhov2}.\\

Recognizing that the competition between two structures is related to the mechanism for the presence of anomalies, a number of effective two length scales models have been proposed \cite{ryzhov1,l-l1,l-l2,l-l3,l-l4,l-l5,hemmer/stell,wilding,camp1,camp2,stanley_scala}. The simplest version of these models is based on the lattice gas structure \cite{l-l6,l-l7,l-l8,l-l9,debenedeti}. In these cases the directionality of the hydrogen bonds is incorporated to the occupation variable with the addition of  an arm variable \cite{debenedeti,l-l4,henriques} and in some cases with the ad hoc increase of volume when the arm values are ordered \cite{l-l2}. Recently a couple of lattice models have been proposed in which no arm variable is presented. Instead of the directionality the anomalies of water are obtained by the competition between two length scales: one attractive and one repulsive \cite{allan, aline}. In both models the presence of attraction is fundamental for the existence of anomalies.

The existence of density and diffusion anomalies in lattice models with no directionality indicates that the anisotropy in this system is not fundamental for the existence of these anomalies. Then the question arises if the attractive interaction is relevant or not for a system to exhibits anomalies. In order to answer to this question, in this paper we study the phase diagram of a triangular lattice gas model that presents only repulsive interactions: a infinity repulsive hard-core interaction with the nearest neighbors and a finite repulsive interaction with the next to nearest neighbors. Our basic assumption is that these two repulsive length scales can reproduce a competition between two ordered structures as presented in water, that we believe is responsible for the anomalies. Hence, we test if this model exhibits the density and diffusion anomalies. In addition we also verify if in the vicinity of the anomalous region multiple liquid phases are presented. Finally we also verify if the criticality is followed by a dynamical fragile-to-strong transition, particularly in the vicinity of the diffusion anomalous region in the chemical potential {\it vs} temperature phase diagram. \\

This article is divided in sections as follow. In section 2 we present the model, in section 3 the Monte Carlo simulations are described. In section 4 results are shown and conclusions are presented in section 5.


\section{The Model}
\label{the_model}
We considered a two dimensional triangular lattice gas model with $L^2$ sites. Each lattice site thus can be empty or occupied by a particle, has six nearest neighbors with a distance $a$ and six next to nearest neighbors with distance $\sqrt{3}a$ where $a$ is the length scale of the model. The particles interact by a two-scale repulsive potential: an infinite repulsive hard core interaction with its nearest neighbors and a finite repulsive interaction $\epsilon$ with its next to nearest neighbors. Therefore, if a site is occupied, its six nearest neighbors must be all empty and this particle will interact with its six next to nearest neighbors if they are occupied too \cite{Lomba}.
Thus, the Hamiltonian for this model can be written as:
\begin{eqnarray}
  {\cal \tilde{H}}= \frac{1}{2} \sum_{i=1}^{L^2} \sum_{\langle i,j \rangle } n_i n_j \epsilon_{ij}  
  \label{hamiltoniana}
\end{eqnarray}
where $\langle i,j \rangle$ indicates the interaction of next to nearest neighbor pairs of sites. If a site $i$ is empty $n_i=0 $, and if it is occupied: $n_i=1$ and $\epsilon_{ij}=\infty$ for nearest neighbors and $\epsilon_{ij}>0$ for the next to nearest neighbors. In the Grand Canonical Ensemble, we can write an effective Hamiltonian, namely
\begin{eqnarray}
  {\cal H} = \sum_{\langle i,j \rangle}^{L^2} \epsilon_{ij} n_i n_j - \mu \sum_{i=1}^{L^2} n_i 
 \end{eqnarray}

And the Grand Potential will be defined as:

\begin{eqnarray}
\Phi(T,\mu)=\langle {\cal H} \rangle -TS 
\label{Phi}
\end{eqnarray}

This model was originally proposed by Almarza {\it et al} \cite{Lomba} and the full phase diagram was obtained by an improved mean field approach and simulations. \\

At $T=0$ it is possible to observe three different structures, depending on the value of the 
chemical potential $\mu$. For negative values of chemical potential, the grand potential is minimal for the empty lattice, so we classified it as a gas phase. For intermediate values of chemical potential, positive but not too high, there is competition between the chemical potential and the free energy. \\

\begin{figure}[!htb]
  \centering
  \subfigure[]{\includegraphics[width=0.25\columnwidth]{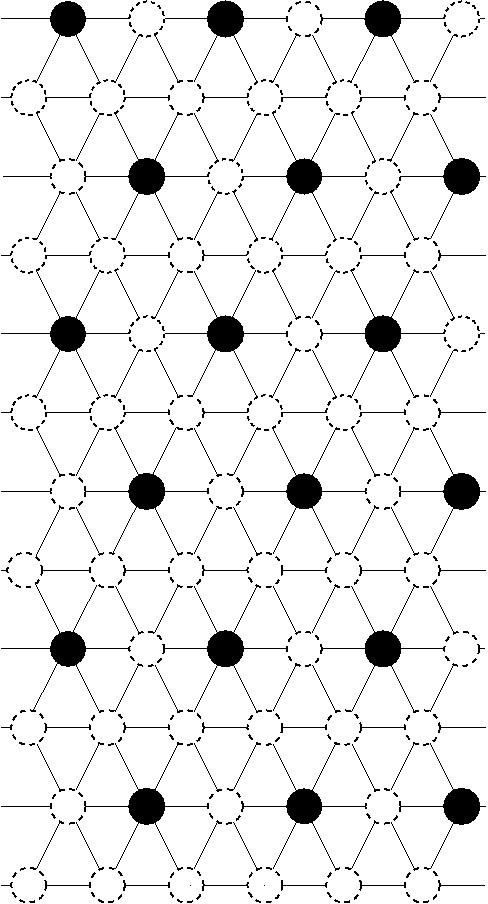}
    \label{T4a}}
  \quad
  \subfigure[]{\includegraphics[width=0.25\columnwidth]{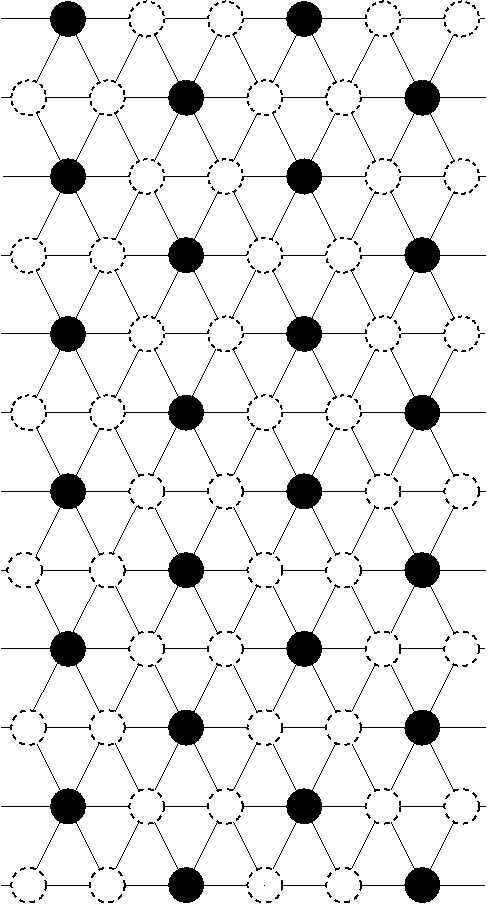}
    \label{T3b}}
    \caption{Representation of structures (a) $T4$ and (b) $T3$. The filled circles represent the sites that are occupied, while the empty circles represents the empty sites. The structure $T4$ has no interactions between next to nearest neighbors particles, because its structure comprehends only the third-nearest neighbors. The structure $T3$ has all next to nearest neighbors occupied, setting it as the maximum occupational configuration for this model.}
 \label{redesT3T4}
\end{figure}

The positive chemical potential tends to fill the lattice, but the energy, that is repulsive, tends to empty it. Due to this competition, for $0 < \mu^* < 12$ ($ \mu^* = \mu / \epsilon $) only the third-nearest neighbors are occupied, generating an organized structure where a quarter of the lattice is occupied and there is no interaction between pairs of next to nearest neighbors, because they do not exist. This phase is called $T4$ \cite{Lomba}. At $\mu^*=0$, the gas and the $T4$ phases coexist at first order phase transition. For high enough chemical potential values, $ \mu^* > 12$, the chemical potential term is more relevant and the system is in its maximum occupation, where a third of the lattice is occupied and each particle has its six next to nearest neighbors occupied, and they interact repulsively. This phase is called $T3$ \cite{Lomba}. At $\mu^*=12$ the $T4$ and $T3$ phases coexist at a first order phase transition. Figure \ref{redesT3T4} illustrates the phases $T4$ and $T3$. \\

As the temperature is increased a new disordered phase appears. In order to understand the behavior of this model for different chemical potential and temperature conditions, we made use of numerical tools, such as Monte Carlo simulations. In the next section we describe the method used to study this model. \\

\section{The Method}
\label{the_method}
In order to obtain the chemical potential {\it vs} temperature phase diagram and the density at all state points, Monte Carlo simulations, particularly the Grand Canonical Monte Carlo Algorithm (GCMC) \cite{Allen} was used. Systems with size $L = 24, 36, 48, 60$ were analyzed. Equilibration times were $t=1 \times 10^7$ Monte Carlo steps. \\

In order to find the order of the transitions the energy fourth order
 cumulant of energy~\cite{salinas_cum} was computed: 
\begin{eqnarray}
V_{\cal \tilde{H}}(L) = 1 - \frac{ \langle ( {\cal \tilde{H}} - \langle {\cal \tilde{H}} \rangle )^4 \rangle_L }{3 \langle ({ \cal \tilde{H}} - \langle {\cal \tilde{H}} \rangle )^2 \rangle_L ^2}\; .
\label{binder_equation}
\end{eqnarray}

This fourth order cumulant identifies a first-order transition for
a negative peak in the $V_L$ value and a continuous transition
for a $V_L$ that vanishes at the criticality. Usually in
this case  the transition is in
between a small positive peak and a small negative
peak.
Considering these tools, it was possible to obtain the chemical potential {\it vs} temperature phase diagram. The Monte Carlo method is considered atemporal, however, in order to analyze the dynamical properties for this model, we considered each Monte Carlo step as a time unit. In order to calculate diffusion in this model using a Monte Carlo simulation, we employed a  two-stages simulation. In the first part, using the grand canonical ensemble, the simulation started with fixed temperature and chemical potential.  It was initialized and iterated until its equilibrium state was reached at $1 \times 10 ^8$ equilibration time. Then, in a second stage, in the NVT ensemble,  we fixed the equilibrated number of particles and we observed how the  particles diffuse  inside the lattice as a function of time (Monte Carlo steps). \\
 
Therefore, for a fixed temperature
and number of particles, at the $NVT$ ensemble, at  each Monte Carlo 
step the following procedure was repeated $N$ times, where $N$ is the number of particles in the lattice. A particle of the $N$ occupied sites
is chosen randomly. Next one of the six nearest neighbors is chosen. If this nearest neighbors has its six nearest neighbors empty (not considering the particle site itself), the particle can move with probability $\exp(-\beta {\cal H})$, otherwise the movement is rejected. Then, for each
time step the $N$ particles can move.\\

For computing the diffusion coefficient, $r(t)$ was calculated by
the average displacement of the $N$ particles at each given time
step $t$ namely
\begin{eqnarray}
  \langle[\Delta r^{i} (t)]^2 \rangle = \frac{1}{N} \sum_{j=1}^N[r_j(t)-r_j(0)]^2
  \label{deltar2}
\end{eqnarray}

We then performed an average of $n$ samples to obtain the
average displacement
\begin{eqnarray}
  \langle[\Delta r (t)]^2 \rangle = \frac{1}{n} \sum_{i=1}^n \langle[\Delta r^{i} (t)]^2 \rangle\; .
  \label{deltar2}
\end{eqnarray}

Considering the Einstein relation to large times,
\begin{eqnarray}
\langle\Delta r (t)^2 \rangle = 4 D t
\label{deltar24Dt}
\end{eqnarray}
it is possible to obtain the translational diffusion coefficient $D$ by measuring the slope of the linear curve of 
$\langle \Delta r (t)^2 \rangle$ as a function of time. The factor $4$ on equation (\ref{deltar24Dt}) refers to a bi-dimensional system. \\

With the diffusion coefficient value it was possible to analyze the diffusion behavior in the chemical potential {\it vs } temperature phase diagram. To observe if there was a fragile-to-strong transition behavior in diffusion for this model we fixed the chemical potential and plotted the diffusion coefficient, obtained by performing an average of $n=1000$ samples, as a function of $T^{*-1}$. \\

\section{The Results}
\label{the_results}
\begin{figure}[!htb]
  \centering
    \subfigure[]{\includegraphics[width=0.47\columnwidth]{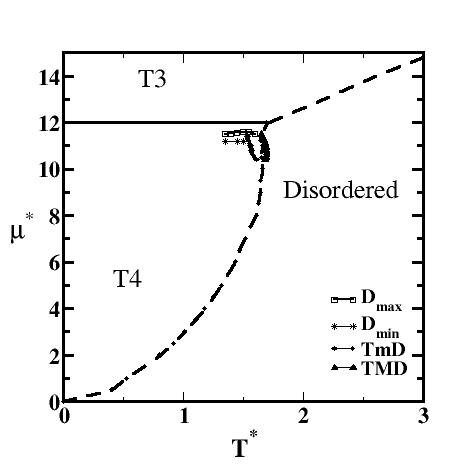} 
 \label{phase1}}
 \quad
  \subfigure[]{\includegraphics[width=0.47\columnwidth]{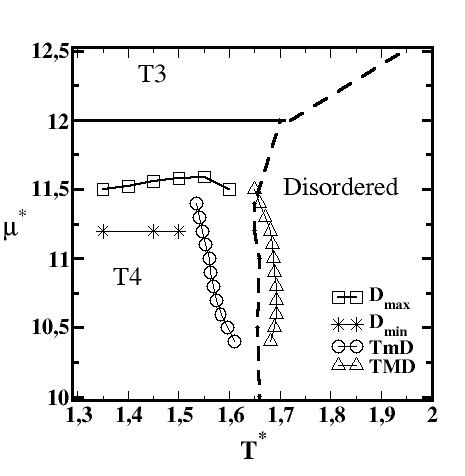}
    \label{phase_zoom}}
 \caption{(a) Reduced chemical potential {\it vs} reduced temperature phase diagram ($T^*=k_BT/\epsilon$ and $\mu^*=\mu/\epsilon$) presenting the coexistence lines for the two ordered phases $T4$ and $T3$ and the disordered fluid phase, besides the region of anomalies where the temperature of maximum (TMD) and minimum (TmD) density lines and the maximum ($D_{max}$) and minimum ($D_{min}$) on diffusion lines were outlined. At $\mu^*=12$ and higher temperature there is a bicritical point in the end of $T4$-$T3$ first order coexistence line. The dashed lines represent continuous phase transition; (b) zoom of anomalous region.}
 \label{phase_diagram}
\end{figure}
The figure \ref{phase_diagram} illustrates the reduced chemical potential {\it vs} reduced temperature $T^*=k_BT/\epsilon$ phase diagram \cite{Lomba} where the different phases are shown. At intermediate chemical potential, $0<\mu^*<12$, as temperature is increased there is a phase transition between the $T4$ and the disordered phase while at high chemical potential, $\mu^*>12$, there is a phase transition between the $T3$ phase and the disordered phase.\\

\begin{figure}[!htb]
  \centering
  \includegraphics[width=0.85\columnwidth]{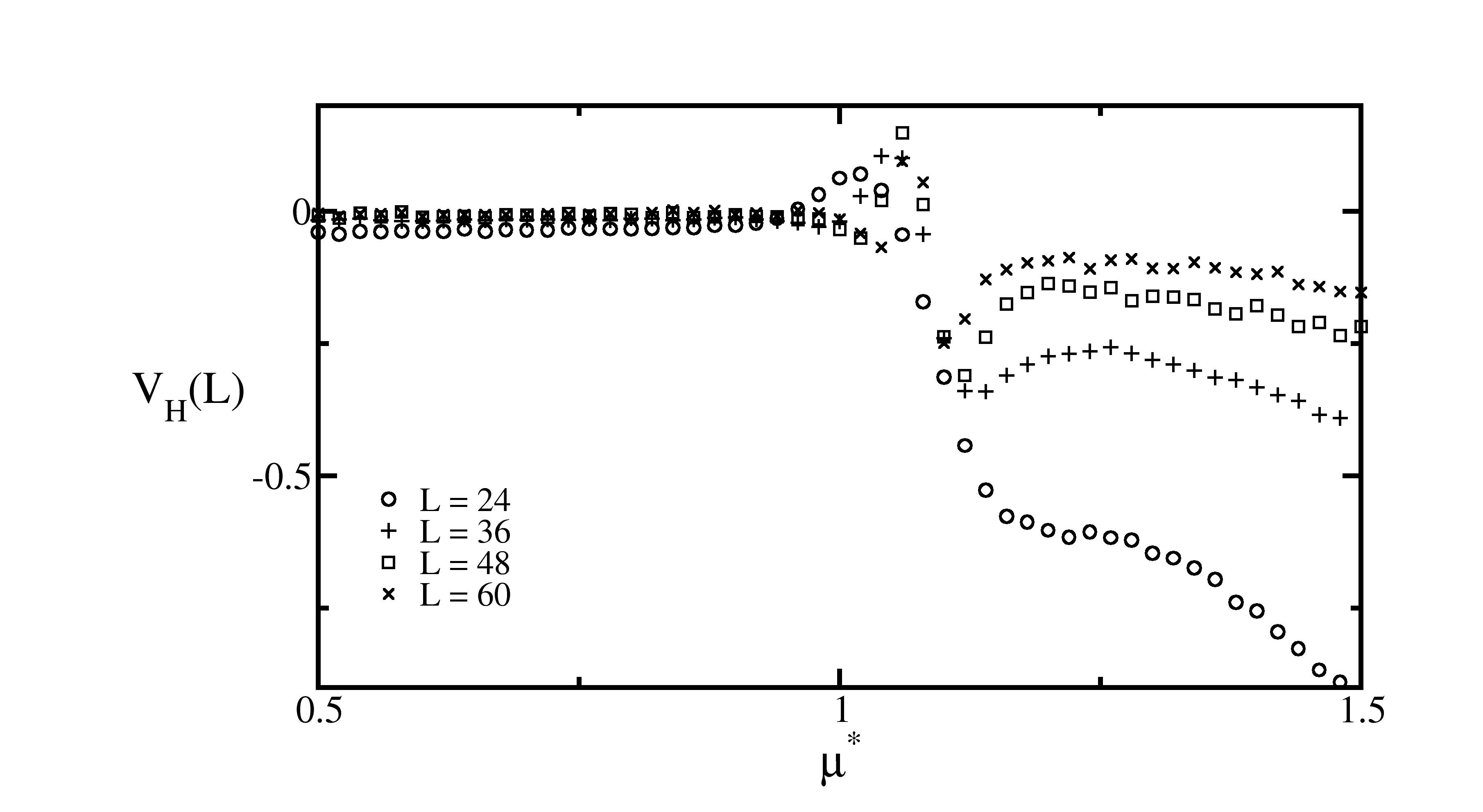} 
 \caption{ Energy fourth order  cumulant {\it vs} chemical potential for $T^* = 0.50$, calculated using equation \ref{binder_equation} for different lattice sizes (L), when it crosses a phase transition line, characterizing a second order phase transition. $T^*=k_BT/\epsilon$ and $\mu^*=\mu/\epsilon$ }
 \label{binder_T050}
\end{figure}
\begin{figure}[!htb]
  \centering
  \includegraphics[width=0.85\columnwidth]{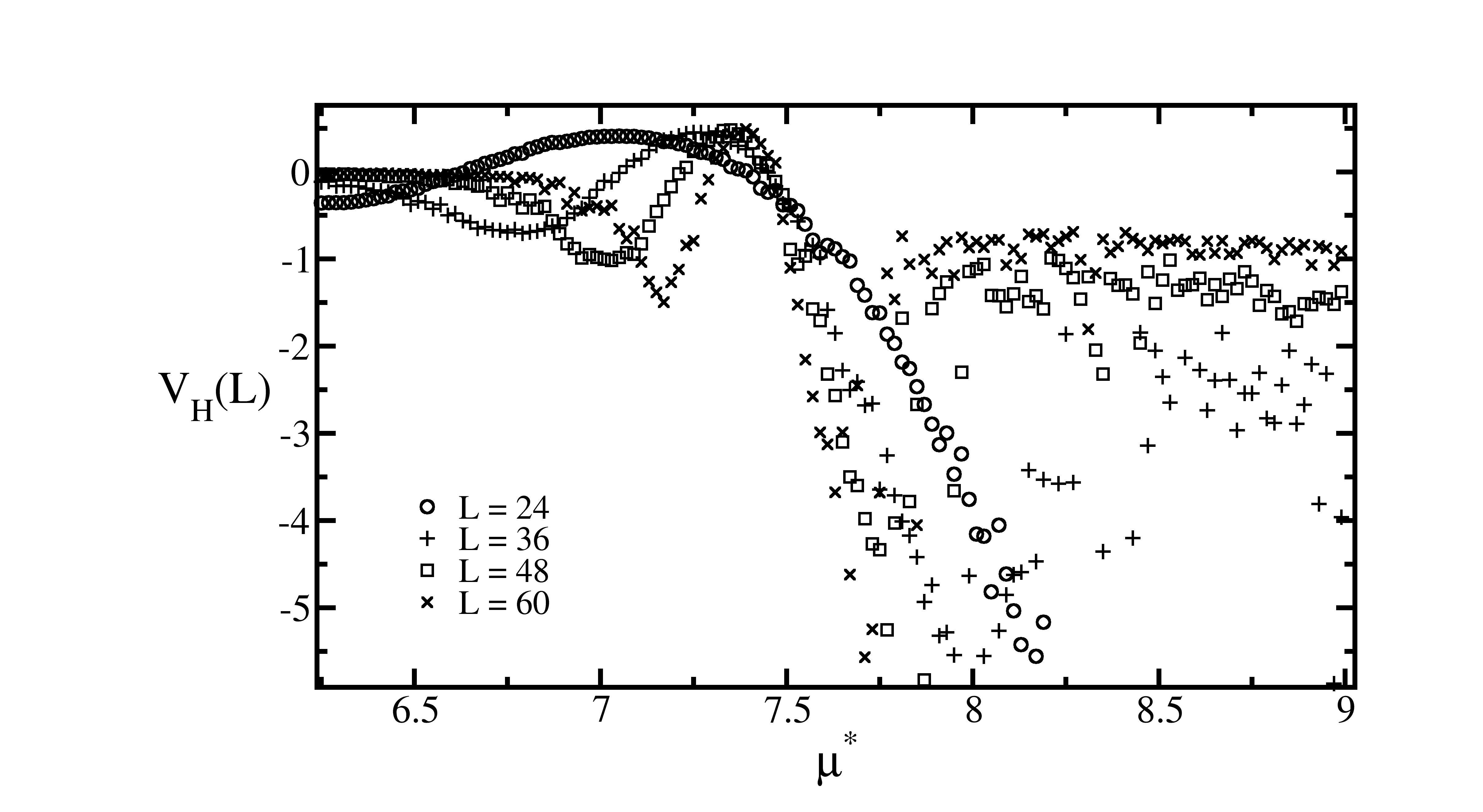} 
 \caption{ Energy fourth order  cumulant {\it vs} chemical potential for $T^* = 1.5$, calculated using equation \ref{binder_equation} for different lattice sizes (L), when it crosses a phase transition line, characterizing a second order phase transition. $T^*=k_BT/\epsilon$ and $\mu^*=\mu/\epsilon$ }
\label{binder_T150}
\end{figure}
\begin{figure}[!htb]
  \centering
  \includegraphics[width=0.85\columnwidth]{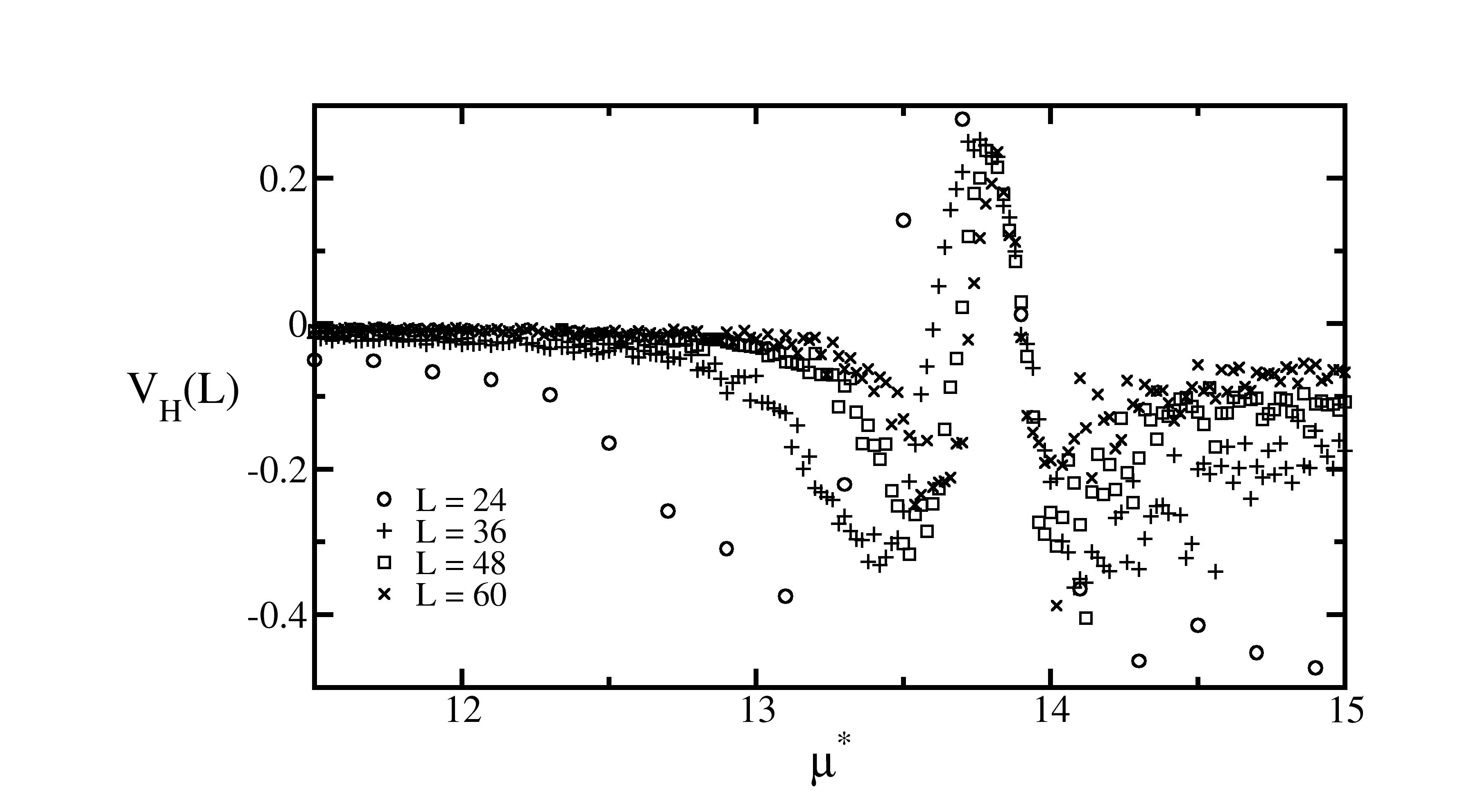} 
 \caption{ Energy fourth order  cumulant {\it vs} chemical potential for $T^* = 2.5$, calculated using equation \ref{binder_equation} for different lattice sizes (L), when it crosses a phase transition line, characterizing a second order phase transition. $T^*=k_BT/\epsilon$ and $\mu^*=\mu/\epsilon$ }
 \label{binder_T250}
\end{figure}

In order to identify the order of these transitions the energy fourth order cumulant 
was  computed.
Figure \ref{binder_T050}, \ref{binder_T150} and \ref {binder_T250} illustrate the behavior of $V_{\cal \tilde{H}}(L)$  for $T^*=0.50$, $T^*=1.50$ and $T^*=2.50$, respectively. For these temperatures the system crosses a second order phase transition from a fluid disordered phase to the ordered phases. The fourth order cumulant show that the $T3$-fluid phase transition line and the $T4$-fluid phase transition line are both continuous transition. Our results
suggest that only at  $T^*=0$ the $T4$-disordered transition becomes first-order. However, at low temperatures other methods
show the existence of a first-order phase transition linked
to the critical line by a tricritical point~\cite{Lomba}. At approximately $\mu^*=12$ and $T^*\leq 1.7$ there is a first-order $T4$ to $T3$ phase transition, as illustrated in figure \ref{phase_diagram}. \\

\begin{figure}[!htb]
  \centering
  \includegraphics[width=0.8\columnwidth]{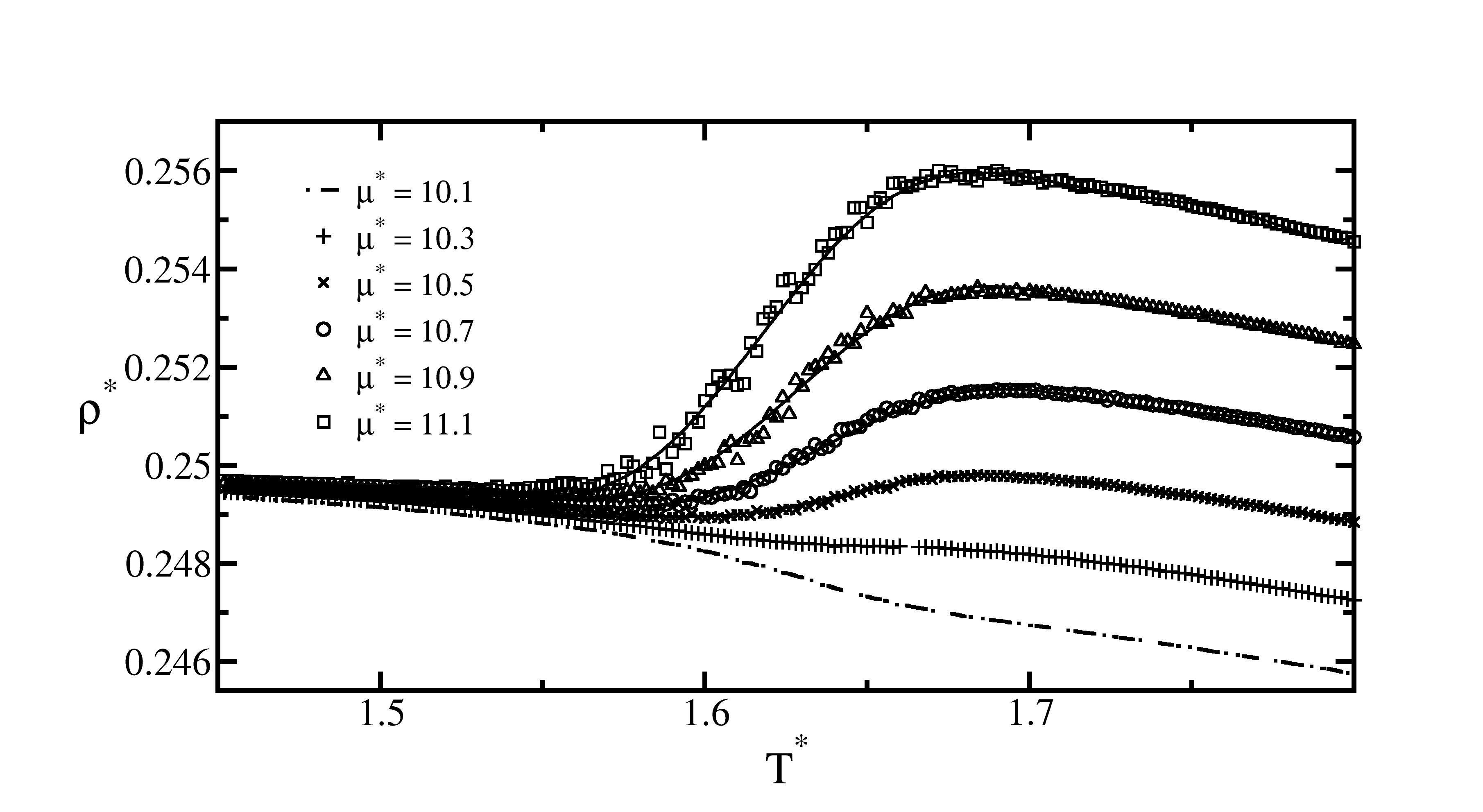} 
 \caption{Plot of density {\it vs} reduced temperature for different values of reduced chemical potential showing a maximum and a minimum on density when we vary the reduced temperature,  $T^*=k_BT/\epsilon$ and $\mu^*=\mu/\epsilon$.}
\label{density_anomaly}
\end{figure}

The density {\it vs} reduced temperature at
constant chemical potential is shown in figure \ref{density_anomaly}. Here we
keep the chemical potential fixed instead of 
the pressure since our simulations are
performed in the Grand Canonical Ensemble. The
temperature of maximum density is the same for 
$\mu$ or $P$ fixed.  Figure \ref{density_anomaly} shows a region of chemical potential where the density increases with the increase of $T^*$, what is called density anomaly \cite{Andressa_proceeding}. The temperature of maximum density line is shown in \ref{phase_zoom}.
 \\

These minimum and maximum on density are characteristics of anomalous fluids at constant pressure. To relate the observed phenomenon at constant chemical potential and the experimental phenomenon, at constant pressure, we used the Gibbs-Duhem relation:
\begin{eqnarray}
  S dT- V dP+ N d\mu = 0
\end{eqnarray}
If we integrate it at constant temperatures we find that 
\begin{eqnarray}
\int^{p_f}_{p_i}dP=\int^{\mu_f}_{\mu_i}\frac{N}{V} d\mu
\end{eqnarray}
Considering the gas phase as the initial one, ($P_i=0$) we have:
\begin{eqnarray}
P = \int^{\mu_f}_{\mu_i}\rho d\mu
\end{eqnarray}
So it is possible to characterize the density anomaly at constant chemical potential as the same anomaly at constant pressure.

\begin{figure}[!htb]
  \centering
  \includegraphics[width=0.8\columnwidth]{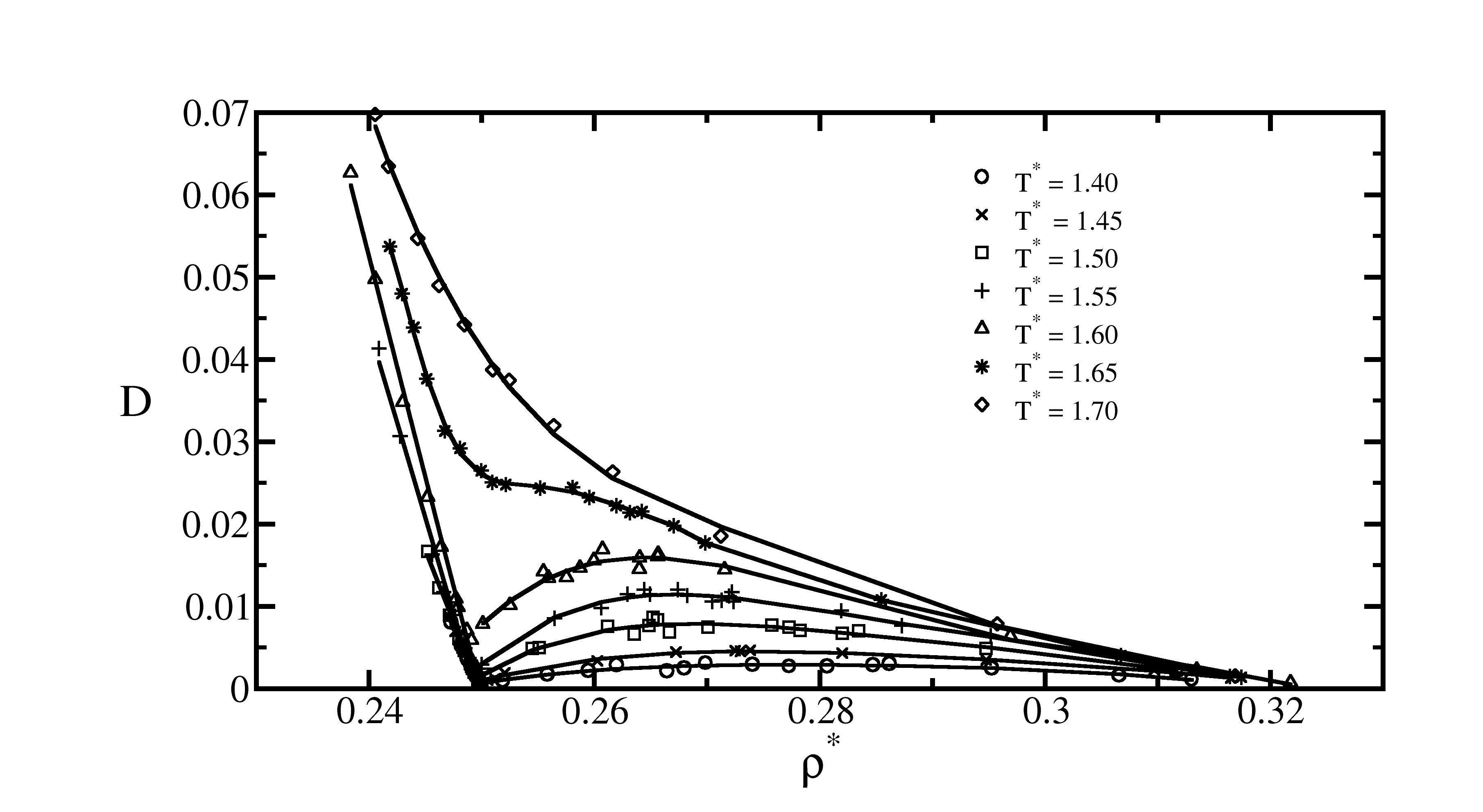} 
  \caption{Plot of diffusion coefficient density {\it vs} density for different values of reduced temperature showing a maximum and a minimum on diffusion when we vary the density,  $T^*=k_BT/\epsilon$. Average of $n=150$ samples.}
\label{diffusion_anomaly}
\end{figure}

The diffusion {\it vs} density for different reduced temperature was also computed and is illustrated in figure \ref{diffusion_anomaly}. For a certain range of temperatures the diffusion coefficient increases with the increase of $\rho^*$. The location in the chemical potential {\it vs} temperature phase diagram of the maximum and minimum of $D$ are shown in figures \ref{phase_diagram} and \ref{phase_zoom}. The anomaly in density and diffusion are characteristics of anomalous liquids as water \cite{Andressa_proceeding}. \\

This model has four types of structures: a gas phase, two ordered and one disordered phase. For high values of chemical potential ($\mu^*>12$) and small temperatures the $T3$ ordered structure is the more stable phase. As the temperature is increased, the system changes from $T3$ to the fluid phase continuously when it crosses the critical line. At approximately $\mu^*=12$, for small temperatures, the $T3$ structure changes discontinuously to the $T4$ structure when the chemical potential is decreased, characterizing a first order phase transition. In the crossover of the $T3$-$T4$ coexistence line and $T3$-fluid critical line there is a bicritical point. Decreasing the chemical potential we can observe two phases: the $T4$, for low temperatures, and fluid, for higher temperatures. The transition from $T4$ to fluid phase is continuous. The transition from $T4$ to fluid phase is continuous. Studies of
Almarza et. al~\cite{Lomba} indicates that at low temperatures the
continuous transition becomes a first-order at a tricritical point. 
Our results with the Fourth Order Cumulant indicate that the transition 
is continuous for $T>0$. It is important to point out that
at such low temperatures our Monte Carlo analysis might be
trapped in a metastable configuration what does not allow
us to see the multicritical point. For $\mu^*<0$ the system is in the gas phase. Near the bicritical point the $T4$-fluid coexistence line is reentrant and we observe some anomalous properties in this region, such as density anomaly and anomaly on self diffusion. \\

\begin{figure}[!htb]
  \centering
  \includegraphics[width=0.8\columnwidth]{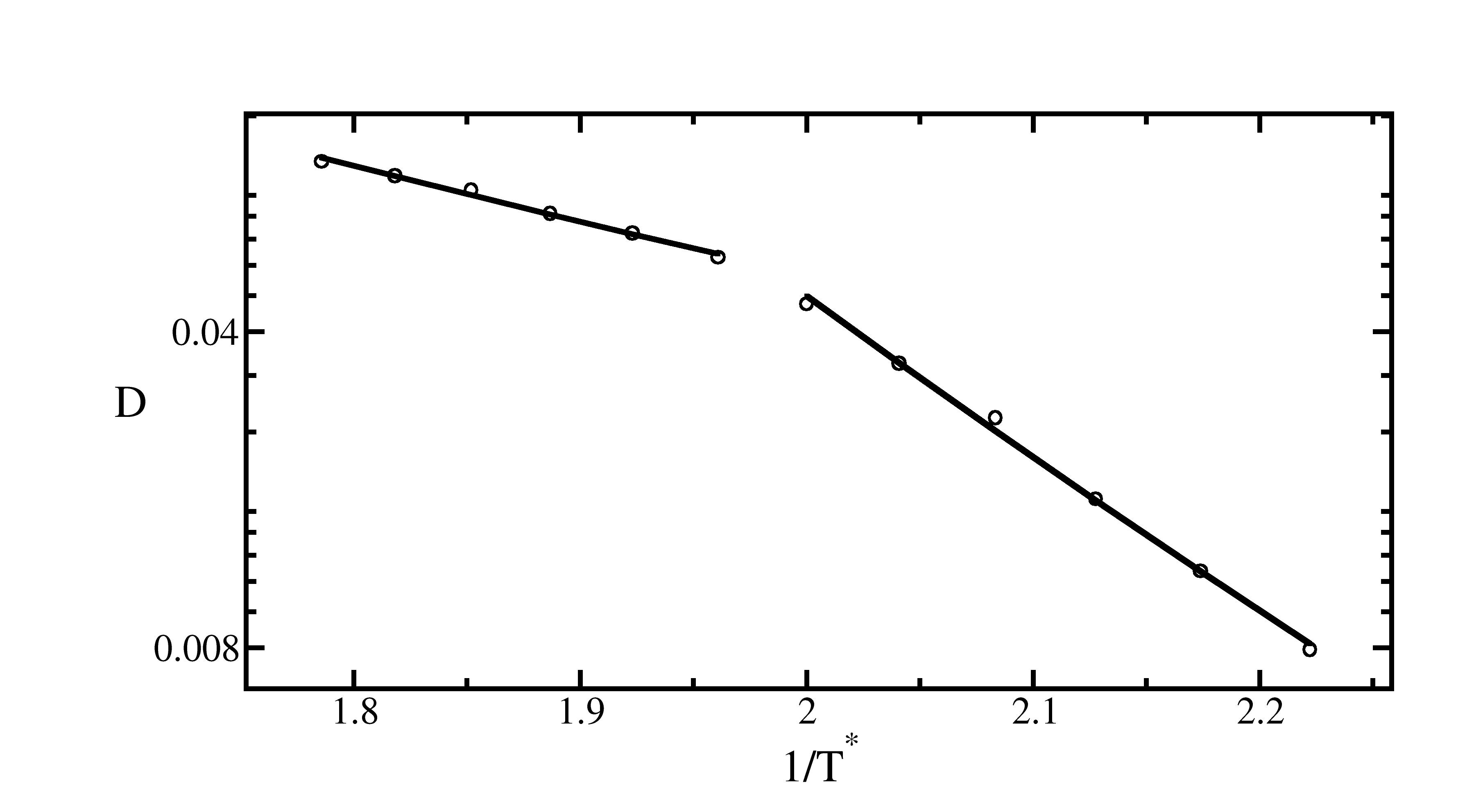}
   \caption{Diffusion coefficient {\it vs} the inverse of temperature in a semi-log plot for $\mu^*=1.00$. For small values of temperature we observe Arrhenius diffusion and for higher values of temperature an Arrhenius diffusion too with a change on curve slope, characterizing a strong-to-strong transition.}
 \label{fst_u100}
\end{figure}

\begin{figure}[!htb]
  \centering
  \includegraphics[width=0.8\columnwidth]{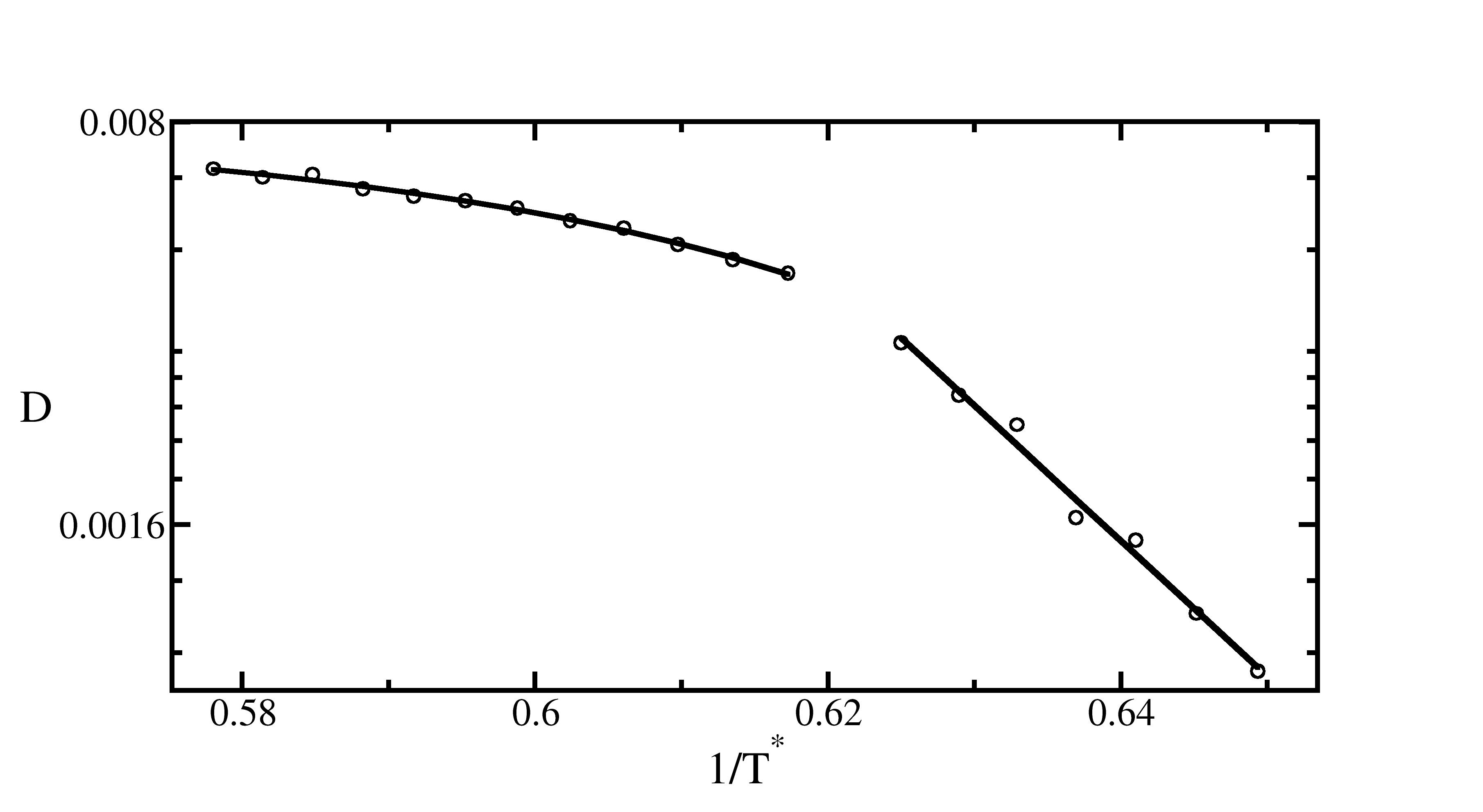}
   \caption{Diffusion coefficient {\it vs} the inverse of temperature in a semi-log plot for $\mu^*=11.55$. For small values of temperature we observe Arrhenius diffusion and for higher values of temperature a non-Arrhenius diffusion, characterizing a fragile-to-strong transition.}
 \label{fst_u1155}
\end{figure}

\begin{figure}[!htb]
  \centering
  \includegraphics[width=0.8\columnwidth]{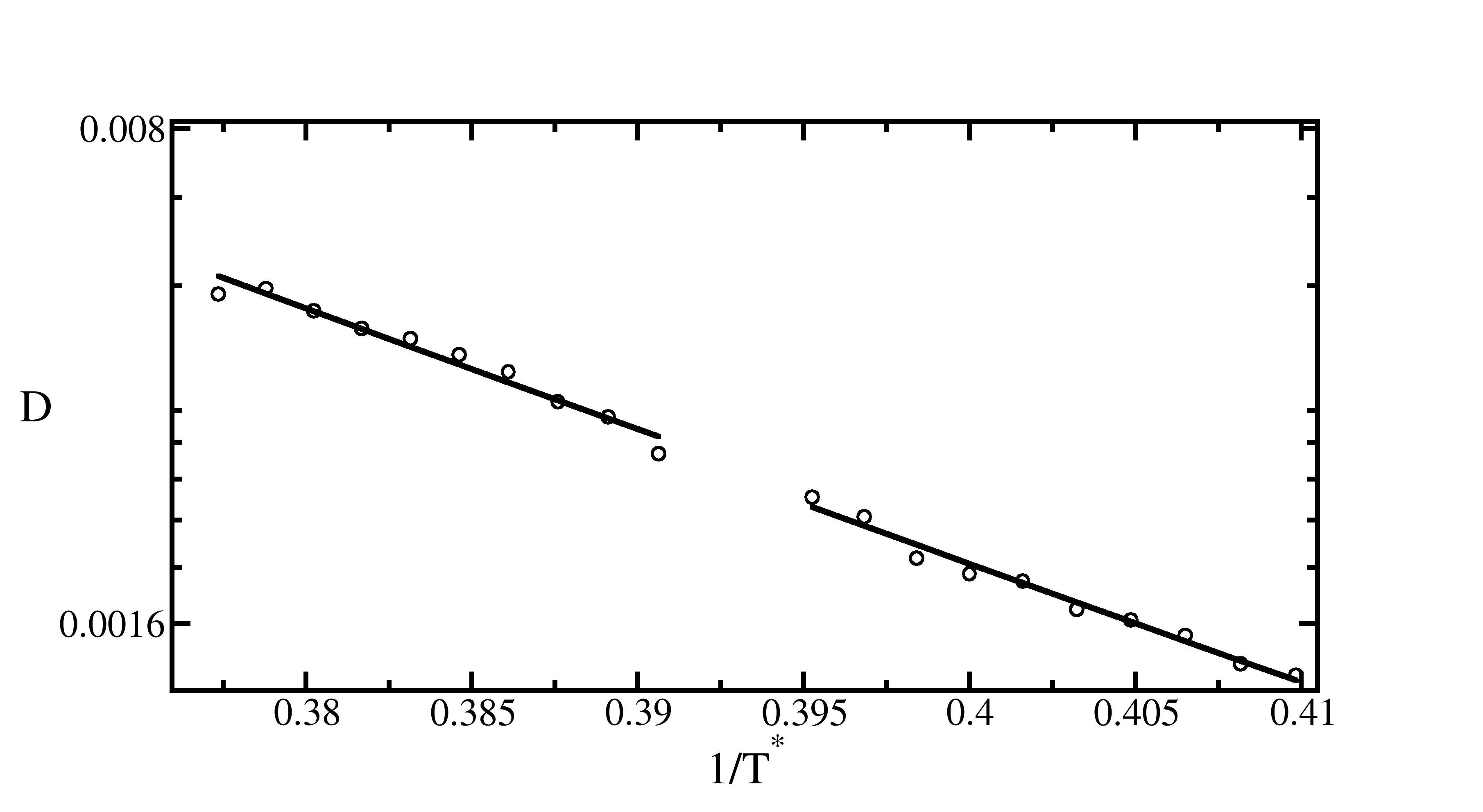}
   \caption{Diffusion coefficient {\it vs} the inverse of temperature in a semi-log plot for $\mu^*=14.00$. }
 \label{fst_u1400}
\end{figure}

In order to understand the relation between the critical lines and a possible fragile-to-strong-transition, the diffusion coefficient as a function of temperature was calculated for a number of reduced chemical potentials. Figure \ref{fst_u1155} illustrates the diffusion coefficient {\it vs} inverse temperature for $\mu^*=11.55$ very close to the region of anomalies. In this region there is a fragile-to-strong transition. In figure \ref{fst_u100}, for $\mu^*=1.00$ a strong-to-strong transition is observed. Figure \ref{fst_u1400} shows the behavior of $D$ for higher values of reduced chemical potential, at $\mu^*=14.00$. In this case no dynamic transition is observed. \\

\section{Conclusions}
In this paper we have shown that a pure repulsive system is able to exhibit density and diffusion anomalies.\\

The chemical potential {\it vs} temperature phase diagram was obtained, showing four phases: a gas phase, two ordered phases with different density values and a disordered fluid phase with variable density. The ordered phases were called $T4$ and $T3$, with fourth occupied sites and a third occupied sites, respectively. The $T4$ phase is less dense and does not has interaction between particles, in the other hand, the $T3$ phase is more dense and has all sites interacting with its six next to nearest neighbors.\\

We observed anomalous behaviors in this model for density, self diffusion and a fragile-to-strong transition. The fragile-to-strong transition was only observed in the anomalous region of phase diagram.\\

Considering that water has a competition of two structures: one with less density and rigid, with all four hydrogen bonds, and another that is more compact but flexible, with less hydrogen bonds, it is possible to conclude that the attractive and directionality of hydrogen bonds is not necessary for the system to be anomalous. The competition between two accessible structures can be the main ingredient for it. In addition we have shown that the fragile-to-strong transition observed in liquid water \cite{fstran,fst2,fst3,fst_stanley} close to the second critical point and usually associated with criticality is also seen in our system. However, in our case it only occurs close to the temperature of maximum on density line. This suggests that the fragile-to-strong transition is not associated to criticality but to the criticality that arises from the competition between ordered structures. Similar behavior was obtained in other lattice models \cite{fst_szortyka}.\\

\section*{Acknowledgment}
We acknowledge Brazilian science agencies CAPES and CNPQ for financial support and Centro de F\'isica Computacional - IF (CFCIF) for computational support. \\








\end{document}